# Explanation of Superluminal Phenomena Based on Wave-Particle Duality and Proposed Optical Experiments

Hai-Long Zhao

**Abstract:** An explanation for superluminal phenomena based on wave-particle duality of photons is suggested. A single photon may be regarded as a wave packet, whose spatial extension is its coherence volume. As a photon propagates as a wave train in vacuum, its velocity is just the speed of light. When it tunnels through a barrier as a particle, its wave function collapses and it will travel faster than light. Superluminal motion can occur only within the coherence length and the time constrained by uncertainty principle. A massive particle cannot be superluminal during the tunneling process. So superluminality does not violate causality. As for the superluminal and negative group velocities in anomalously dispersive medium, they are merely reshaping effect of the pulse, and they will become subluminal at large distances. A couple of experiments are proposed to test the superluminal phenomena.

## 1 Introduction

Since the birth of special relativity, the speed of light has been considered as the highest speed of massive objects in the universe. According to the mass-energy relation $E = m_0 c^2 / \sqrt{1 - v^2 / c^2}$, the speed of an object cannot reach or exceed the speed of light. If $v \geq c$, the mass-energy relation will be invalid. Only massless particles can travel at the speed of light. If superluminal phenomena did exist, one could go back to the past and killed his grandfather, which is obviously a paradox. Then it seems that superluminal motion is impossible.

The above argument is based on the classical theory. Microscopic particles have many characteristics different from those of classical particles, and their behaviors cannot be predicted with common sense. The speed of light we have measured so far is only the speed of photons propagating in the state of wave in vacuum. We know that microscopic particles possess the property of wave-particle duality. When photons interact with other particles or tunnel through a barrier, they behave as particles. The velocity of photons interacting with other particles in the state of particles (such as the emission and absorption process of photons) is hard to obtain, but the tunneling velocity of photons can be measured experimentally. At present, many experiments have demonstrated that the velocity of photons tunneling through the barrier is faster than $c$. These experiments include barrier tunneling of photons [1,2], microwave tunneling through undersized waveguide [3,4], superluminal experiments of microwave in open space [5-8], microwave traversing double-prism [9-11]. Besides these tunneling experiments, there are also experiments where light or electromagnetic pulses pass through anomalous dispersion region [12-16]. For a review of superluminal issue, one may see [17].

The superluminal phenomena in barrier tunneling are usually explained with evanescent waves, which are accelerated through potential barriers and therefore can travel faster than light. In the following, we suggest a natural explanation for barrier tunneling process based on wave-particle duality. On the other hand, some theories and experiments indicated that group velocity may be negative under certain conditions [12-16]. We show that it is only an illusion arising from improper

E-mail: zhlzyj@126.com

measurement method. In addition, several experiments are proposed to verify superluminal phenomena.

## 2 Explanation of superluminal phenomena

### 2.1 The superluminal tunneling of photons through a barrier

We know that the energy of a photon is $E = \hbar\omega$, and the wavelength is $\lambda = 2\pi c/\omega$ with $\omega$ the angular frequency. In fact, a monochromatic photon with an exact angular frequency $\omega$ does not exist in terms of uncertainty principle. For a photon with energy $E = \hbar\omega$, we should regard it as expectation and its frequency fluctuates around $\omega$. Accordingly, a photon should be viewed as a wave packet instead of a point-like particle. Then how large is its spatial extension? It cannot be the scale of only one wavelength; otherwise we can hardly understand the single-photon interference experiment. As interference can only occur when the optical path difference is smaller than the coherence length, we might as well take the coherence length as the length of the wave packet. For a pulse, its coherence length is $l = \lambda^2/\Delta\lambda$ with $\Delta\lambda$ the spectral width. For a single photon, $\Delta\lambda$ should be regarded as the uncertainty of the wavelength $\lambda$. The coherence length of a photon may be several to thousands times larger than its wavelength. In order to better understand, we may take the wave packet as a train, and the carriage is similar to one wavelength. The whole spatial extension of a photon is its coherence volume.

In order to have an intuitive understanding of the superluminal phenomenon of photons passing through the barrier, let's see an example of a daily activity. If a person walks with a step length of $s$ and a frequency of $f$, then the walking speed is $v = sf$. Assume that there is an obstacle with the height of $H_0$ and the width of $a$ on the road, and the height that the person raises legs up is $H$. If $H \geq H_0$ and $s > a$, the person can pass in most cases. In special cases, the person may stumble and fall. This is equivalent to the case where a beam of light is incident upon a medium, a fraction of the light will pass through and the other be reflected. Now let $H < H_0$, then the person cannot pass through the barrier. However, the person can still find ways to pass. For example, by jumping, as long as the height of the jump is greater than $H_0$ and the distance of the jump is greater than $a$.

Let the energy of the photon be $E$, and the barrier height be $V_0$, $E < V_0$. By analogy with the above example, the photon must borrow at least energy $V_0 - E$ to pass through the barrier (which is similar to the situation where a person needs to prepare considerable strength or energy before jumping). Then we must put a limit on the maximum distance that a photon tunnels at one time to avoid unreasonable situation where a photon can tunnel any distance. In fact, when we use the Schrödinger equation to solve the barrier tunneling problem, the prerequisite is that tunneling can occur only within the coherence length, which can be seen from the wave function inside the barrier

$$\psi = Ae^{ipx/\hbar} + Be^{-ipx/\hbar} \tag{1}$$

which indicates that the incoming wave $e^{ipx/\hbar}$ and the reflected wave $e^{-ipx/\hbar}$ inside the barrier are coherent; otherwise the coherent superposition method cannot be adopted. In order to ensure that the incoming wave and reflected wave are coherent in the whole barrier region, the coherence length must be greater than or equal to the barrier thickness, i.e., $l \geq a$. For an ideal plane wave, the coherence length is infinite, so we do not need to consider the problem of coherence in the barrier tunneling process. For a photon or light pulse whose coherence length is limited, the coherence becomes a problem highlighted.

Someone might wonder what happens in the tunneling process. Return to the example of the person flying over the obstacle. It can be seen that the original step length and pace have lost their meanings. Similarly, we think that as a photon traverses the barrier, it behaves like a particle instead of a wave, that is, the wave function of the photon collapses. $c$ is the speed when photons propagate in the state of wave. When in the state of particle, the photon can travel faster than $c$, just like the jumping speed of the person can be greater than the walking speed. In addition, when light tunnels through the barrier, its phase is constant, which just explains the fact that the phase of evanescent wave is constant in the process of propagation.

When a particle tunnels through a barrier, it needs to increase its energy. We may think that the change of the energy $\Delta E$ is the consequence of the energy fluctuation of the particle, or we may think this energy is borrowed from the external environment, as long as it is returned within the time $\Delta t$ constrained by the uncertainty principle, that is

$$\Delta t \leq \frac{\hbar}{\Delta E} \tag{2}$$

The minimum of $\Delta E$ is $\Delta E_{\min} = V_0 - E$, so the maximum of $\Delta t$ is

$$\Delta t_{\max} = \frac{\hbar}{V_0 - E} \tag{3}$$

which is the saturation effect of barrier tunneling. Hartman proved in 1962 that there is a limit for $\Delta t$ when $a$ increases gradually, and the limit is [18]

$$\tau = \frac{\hbar}{\sqrt{E_0(V_0 - E_0)}} \tag{4}$$

Hartman used a Gaussian wave packet, where $E_0$ is the average energy of the wave packet. So his result of (4) is different from (3), which applies to a plane wave. Both the two situations will lead to saturation effect of tunneling time.

Someone might think that when $a$ tends to infinity while $\tau$ or $\Delta t_{\max}$ remains unchanged, isn't it possible to achieve superluminal communication over a long distance? There are two reasons why this will not happen. Firstly, $a$ cannot be greater than half the coherence length. Secondly, the speed of a few of the photons is not equivalent to the speed of the optical signal (information). An optical signal consists of a large number of photons with different frequencies. Only a few of the photons can tunnel through the barrier, so the signal will be attenuated. When $a$ is very large, the tunneling probability decreases exponentially with $a$, so the signal attenuation is very serious. On the other hand, the probabilities of the photons with different frequencies tunneling through the barrier are different, which will lead to distortion of the signal. Attenuated and distorted signal does not represent the real signal, so the superluminal motion of a few of the photons does not violate causality.

The above is the theoretical hypothesis of superluminality. Now let's see the experimental results. In order to realize superluminal motion, photons must pass through a potential barrier, so the key to the problem is to select appropriate materials to form a potential barrier. At present, there are mainly three types of barriers. The first type is multilayer dielectrics (photonic crystals). When photons propagate in periodic potential in the state of wave, they must satisfy Kronig-Penney relationship. For photons with specific frequencies, this relationship cannot be satisfied, thus the photonic crystal will become a band gap for the photons. Photons can only tunnel through the barrier in particle state. A famous experiment is the two-photon race experiment carried out by R. Chiao et al [1]. A pair of photons are produced simultaneously by the down-conversion of nonlinear crystal, one of them propagates in the air, and the other passes through a photonic crystal with the thickness of about 1um. The measured tunneling velocity of the single photon is $1.7c$. Spielman et al. adopted a similar method, but they used extremely short laser pulse (12fs) to carry out the experiment. Their experiment not only obtained the superluminal results, but also verified the Hartman effect [2].

The second type of barrier is cut-off waveguide, which is equivalent to a rectangular barrier. The barrier height is the photon energy corresponding to the cut-off frequency, and the barrier width is the length of the waveguide. Photons can not pass through the cut-off waveguide as guided wave, but some photons can tunnel through it as particles. The typical experiments are those of Nimtz [3,4], who measured the time for microwave pulse to pass through 114.2mm-long cut-off waveguide to be 81ps, while the time for light to pass through the same length in vacuum is 380ps, so the tunneling speed is $4.7c$. Eqs. (3) and (4) give the time limit for a photon to tunnel through the barrier. Now we test this hypothesis. In Nimtz's experiments, the cut-off frequency of the waveguide is 9.49GHz, the pulse frequency is 8.2~9.2GHz, and the center frequency is 8.7GHz. For simplicity, we take the center frequency as the photon frequency. From Eq. (3) we obtain the up limit of tunneling time of 201ps. While the up limit is 61ps for Eq. (4). The measured time for the microwave pulse to pass through the 100mm-long cut-off waveguide is 130ps, and that through the 114.2mm-long cut-off waveguide is 81ps. It can be seen that for the situation of cutoff waveguides, the result of (3) is more reasonable.

We have supposed that barrier tunneling can occur only within half of the coherence length. Now let's test this hypothesis. In R. Chiao's the two-photon race experiment, the coherence time is about 20fs, so the coherence length is about 6um, while the barrier thickness is 1.1um. In Nimtz's experiments, the center frequency of the microwave is 8.7GHz, the frequency range of the microwave pulse is 8.2~9.2GHz, so the coherence length of the microwave is 300 mm. The maximum length of the cut-off waveguide used in the experiment is 114.2 mm. Both the coherence lengths in the two experiments are larger than the barrier thickness.

The third type of barrier consists of two media separated by an air gap. This type of experiments date back to Bose's microwave tunneling through double-prism experiment [19]. The subsequent frustrated total internal reflection experiments are similar to that of Bose, so we discuss Bose's experiment in detail. The schematic diagram of the experiment is shown in Fig. 1.

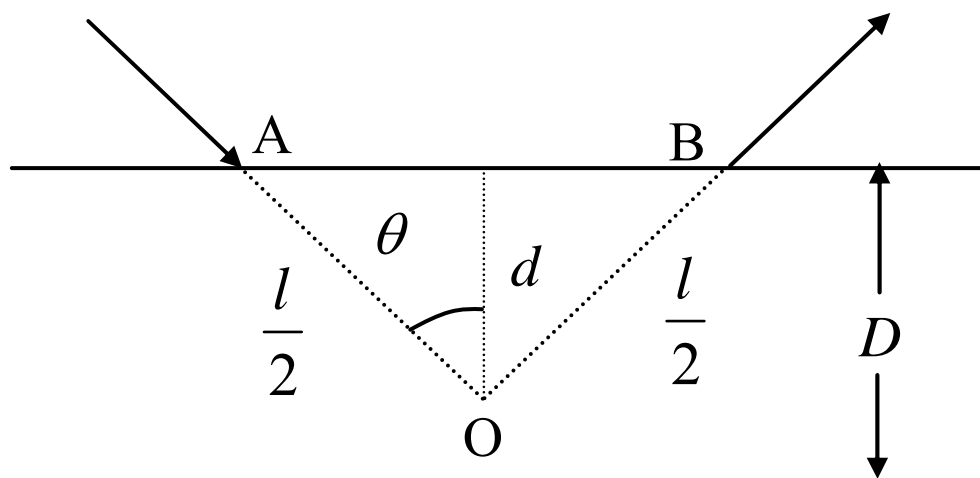

Fig. 1 Illustration of microwave tunneling through double-prism.

We may understand the tunneling process as follows. When the angle of incidence $\theta$ is greater than the critical angle, the total reflection will take place. But a Goos-Hänchen transversal shift appears for the reflected ray, i.e. the incident position is at A while the reflected position is at B. In order to explain this phenomenon, we suppose the turning point of the incident ray is at point O. It can be seen that the wave packet of the photons actually enter the air gap and their distributions are within AOB, whose length is the coherence length $l$. If the entering depth $d$ is smaller than the thickness of the air gap $D$, the photons will be totally reflected at point O. If $d \geq D$, the wave packets of the photons will reach the second prism, then a small fraction of light will be induced out and propagate in the second prism. It follows from Fig. 1 that

$$l = \frac{2d}{\cos\theta} \tag{5}$$

Now we analyze the experimental data of Bose. Table 2-1 is the relationship between the angle of incidence and the minimum thickness of the air gap required to generate total reflection. Table 2 is the coherence length calculated according to the data in Table 1. The critical angle in the experiment is 29°.

Table 1 Relationship between angle of incidence and the minimum thickness of the air gap

| $\theta$ (°) | $d$ (mm) |
|---|---|
| 30 | 13~14 |
| 45 | 9.9~10.3 |
| 60 | 7.2~7.6 |

Table 2 The coherence length calculated according to Table 1

| $\theta$ (°) | $l$ (mm) |
|---|---|
| 30 | 30.02~32.33 |
| 45 | 28.00~29.13 |
| 60 | 28.80~30.40 |

It can be seen from Table 2 that the calculated $l$ is 28~32mm. Taking into account the sensitivity of the measuring device and the experimental uncertainties, we may think that the data in Table 2 agree with Eq. (5). Bose's experiment also indicated that when the air gap is thick sufficiently for total reflection, a portion of the microwave will still be induced out if a thin piece of cardboard or any other refracting substance is inserted into the air gap, which shows that there exist twice tunneling processes for the incident microwave, the first from the first prism to the inserted refracting substance, and the second from the refracting substance to the second prism. A similar experiment has been done by Mugnai [20], where the microwave is incident upon a diffraction grating made of metal strips while it is induced out by a paraffin prism. This situation is similar to the case where there is a stream. When the stream is not wide, a person can jump across it. When the stream is wide enough, a stone may be placed in the middle of it, and the person can step on the stone and jump over it twice. If there exist several tunneling processes, the whole traversal velocity may be slower than $c$. For example, in the experiment of Nimtz [21], an undersized waveguide was filled with carbon loaded urethane foam. During the tunneling process, a photon may interact with a couple of electrons of the foam. Due to the time delay arising from the absorption-emission process, the tunneling velocity of the photon will slow down. In Nimtz's experiment, a traversal velocity of $0.7c$ was obtained.

Besides the above experiments, the superluminal propagation of microwave near the transmitting antennae also belongs to the third type of barrier tunneling, where the air between the transmitting and receiving antennae forms a barrier. When the distance between the two antennae is smaller than half of the coherence length of the microwave, there exists microwave tunneling from the transmitting antenna to the receiving antenna, and the tunneled wave is called evanescent wave. If the distance is larger than half of the coherence length, superluminal propagation disappears. It should be noted that there exist synchronously evanescent wave and radiant component near the transmitting antenna. The radiant component must be suppressed in order to make superluminal phenomenon obvious. In the experimental setup of Giakos and Ishii [22], mis-alignment of the receiving antenna makes the evanescent wave component dominated, as shown in Fig. 2. In the case of Fig. 2(b), the receiving antenna can receive more tunneled microwave signal, so the superluminal phenomenon is more obvious. In the meanwhile, the distance between the two antenna walls is smaller compared to the situation of Fig. 2(a), so superluminal phenomenon can be observed at a larger distance of $d$, just as demonstrated by the experimental results. In fact, when the two horn antennae are placed face to face, there will also exist tunneling of evanescent wave. But in this case the radiant component dominates, so superluminal phenomenon is too weak to be observed.

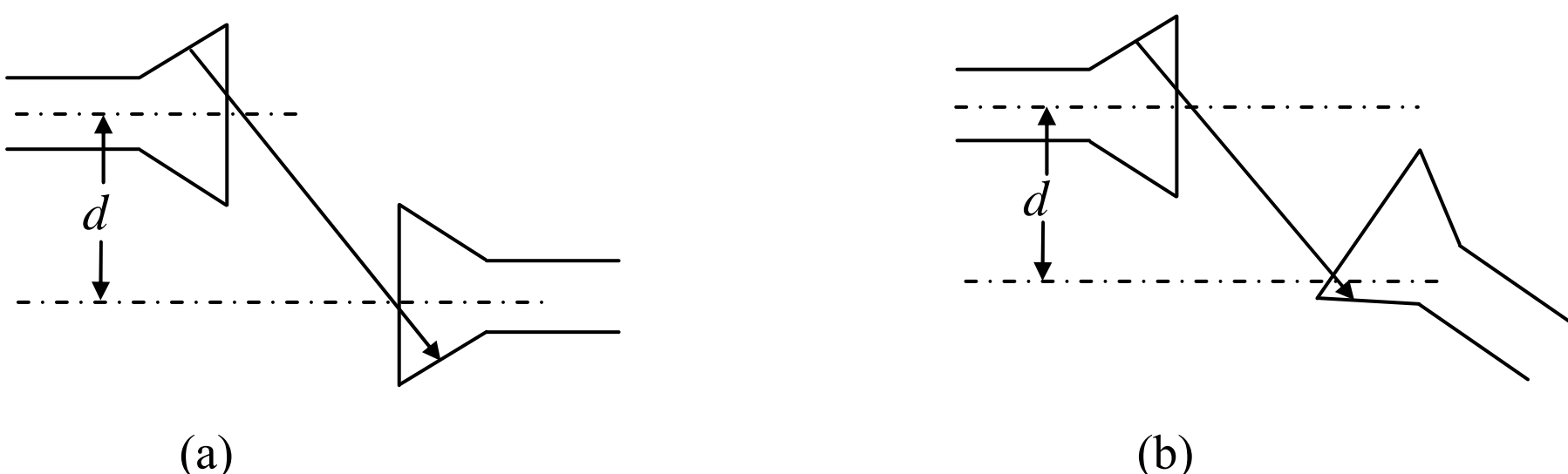

Fig. 2 Giakos-Ishii's experimental setup.

A similar experiment was carried out by Mugnai [23], where the evanescent wave is first induced out by a reflecting mirror and then received by a receiving antenna, as shown in Fig. 3.

There are twice tunneling processes here, one is from the transmitting antenna to the mirror and the other from the mirror to the receiving antenna. Due to the time delay between the two tunneling processes, the superluminal phenomenon is not obvious, and only a speed of $1.053c$ is observed.

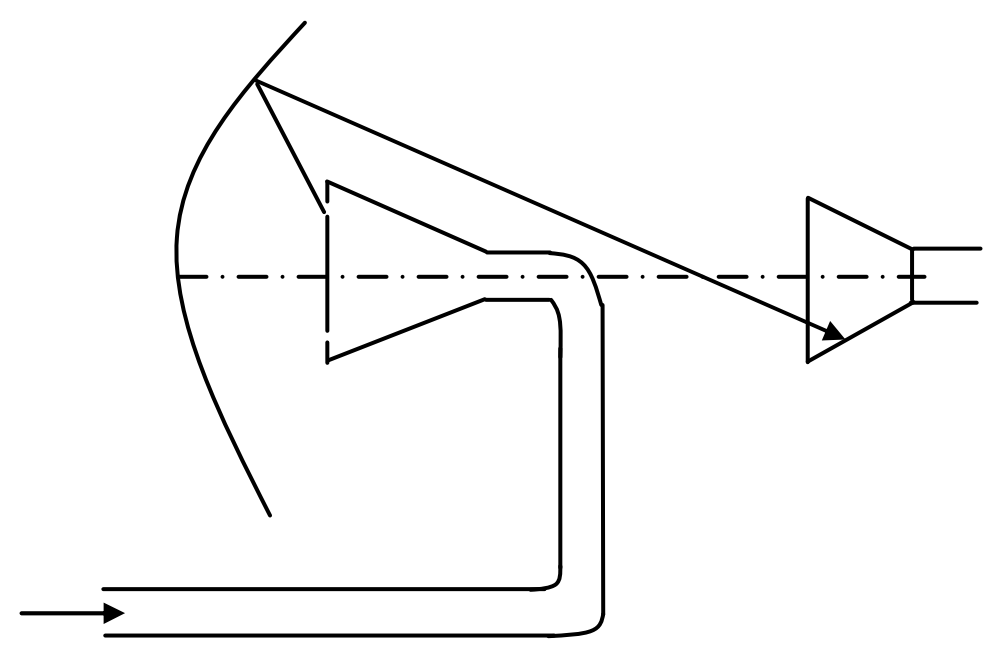

Fig. 3 The experimental setup of Mugnai et al.

If we try to understand the above experimental phenomena according to classical electromagnetic theory, it can be regarded that there exist oscillating standing waves near the transmitting antenna, or we start with the electromagnetic fields of dipole antenna [24]

$$\mathbf{H} = \frac{ck^2}{4\pi}(\mathbf{n}\times\mathbf{p})\frac{e^{ikr}}{r}(1-\frac{1}{ikr}) \tag{6}$$

$$\mathbf{E} = \frac{1}{4\pi\varepsilon_0}\left\{k^2(\mathbf{n}\times\mathbf{p})\times\mathbf{n}\frac{e^{ikr}}{r}+[3\mathbf{n}(\mathbf{n}\cdot\mathbf{p})-\mathbf{p}](\frac{1}{r^3}-\frac{ik}{r^2})e^{ikr}\right\} \tag{7}$$

The terms in the above equations containing $1/r$ are radiant components. It can be seen that there is energy of electromagnetic fields stored near the antenna besides the energy radiated. In the general case, the stored energy does not radiate outwards and there exists only electromagnetic energy current loop near the antenna. If there is a conductor or other substances within half of the coherence length of the electromagnetic pulse, the stored energy will be induced out. The principle is the same as that of the energy exchange in fiber coupler. The only difference is that the coherence length of the microwave is much larger than that of the light, so energy exchange effect can be observed within a larger distance. Because the energy exchange is realized by evanescent wave, superluminal phenomenon can only be observed near the transmitting antenna.

We then discuss the barrier tunneling of a particle with non-vanishing mass. Suppose its initial speed is $v_0$. If the tunneling velocity $v$ could reach $c$, then the energy it borrows from external circumstance $\Delta E = m_0c^2/\sqrt{1-(v/c)^2} - m_0c^2/\sqrt{1-(v_0/c)^2}$ would be infinite, which is impossible. So the traversal velocity cannot reach $c$, and the expression $m = m_0/\sqrt{1-(v/c)^2}$ will never be imaginary. Now we see the experimental results. In the experiment of Ref. [25], the semiclassical traversal time in Josephson junction was measured to be the order of 100 ps. Suppose the thickness of the Josephson junction to be 10nm. Then the tunneling velocity of the electrons is about 100m/s. In the experiment of [26], the traversal time for electron tunneling water was measured. For distances of the order of 1nm the tunneling times computed with Büttiker-Landauer approach are in the range of 0.1~1fs. Then the tunneling velocity of the electrons is about $10^6 \sim 10^7$ m/s. In a recent experiment [27], The Rb atoms in the Bose-Einstein condensation state tunnel through a 1.3um-thick optical fiber, and the measured time is 0.61ms at the lowest energy for which tunneling is observable, then the tunneling speed is $2.1\times10^{-3}$m/s.

### 2.2 The superluminal velocity and negative group velocity in anomalous dispersion region

Apart from the superluminal motion arising from barrier tunneling of photons, there is also the propagation of light in the anomalously dispersive medium, which is often mistaken for superluminal velocity, or even negative group velocity. In order to understand this abnormal phenomenon, we first see the propagation process of light in the medium from the microscopic perspective. When light propagates in the medium, photons will interact with the bound electrons in the medium, and the electrons are forced to oscillate under the interaction of light wave and emit secondary wave. The two together constitute the electromagnetic wave propagating in the medium. Because this effect retards the propagation of light, the speed of light in the medium will decrease, just as the speed of wind will decrease when passing through the forest. Only a few of the photons that do not interact with the electrons can travel with the speed of $c$, and these photons form the wave front. As there are large numbers of electrons in the medium, the probability that the photons do not interact with any electron is the smallest and so is the amplitude of the wave front. The second smallest amplitude is the photons that interact with only one electron and the next is the photons that interact with two electrons……. Certainly, the probability that the photons interact with all the electrons is also the smallest.

As a light or electromagnetic pulse comprises large numbers of photons, we must take into account the shape deformation arising from the different propagation velocities of the photons with different frequencies, i.e. reshaping effect, which will result in the delay shift of the peak of the pulse for normally dispersive medium, and advancement shift for anomalously dispersive medium. We may explain intuitively the latter effect as follows. The photons with higher frequency have faster propagation velocity in anomalously dispersive medium, while those with lower frequency travel slowly. As the photons with higher frequency distribute mostly in the rising and falling edges of the pulse, the rising and falling edges of the pulse will shift forward compared to the original pulse. In this case, the shape of the pulse changes; while the velocity of the wave front is still $c$, as shown in Fig. 4.

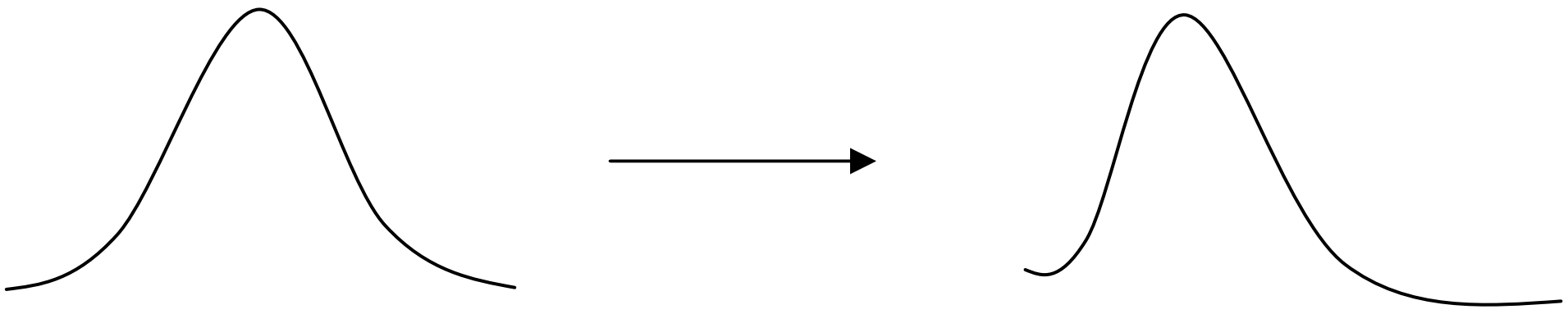

Fig. 4 The shape deformation of the pulse propagating in anomalously dispersive medium.

If that were the whole story, it would be easy to understand the superluminal motion of pulse in abnormally dispersive medium. What is difficult to understand is the so-called negative group velocity, which does not mean the reverse velocity of the wave, but the velocity greater than infinity. This concept is first proposed by Brillouin. In 1914, when he solved Sommerfeld integrals with saddle-point method, he derived a curve, by which he found that at the central frequency of the signal, $c/v_g$ could be 0 (which means $v_g = \infty$) and then turned into a negative value ($v_g < 0$). Therefore, he concluded that the group velocity $v_g$ can be greater than the speed of light, be infinite, or be even negative [28]. Of all the experiments on negative group velocity, the one carried out by Wang et al. in 2000 is the most famous. The experiment found that "the peak of the pulse appears to leave the cell before entering it" [12]. According to the general understanding, the speed of light in the medium is at most infinite. How can there be a negative group velocity? The confusion of the experiment is that it takes 0.2 ns for the laser pulse to pass through the 6cm-long caesium vapour cell, while the pulse advancement shift is 62ns. Besides this experiment, many authors also claimed to achieve negative group velocity experimentally. To explain this paradox, let's see a similar macroscopic example. Suppose a giraffe comes to a stream and stretches its neck to the opposite bank. As seen by the ants at the bank, the giraffe's head has flown across the stream but its legs and feet still stay where they are. In the experiment of Wang et al., the pulse width (pulse peak half width) is 3.7us, and the whole pulse duration is about 8 us. Compared with the pulse width of 8us, the advancement shift of the peak of the 62ns is small, but is very large compared with the 0.2ns required for the peak of the laser pulse to pass through the 6cm-long vapour cell, which brings about the illusion of negative group velocity. In fact, what is observed in the experiment is the advancement shift of the peak of the pulse, not the time that the rising or falling edge of the pulse passing through the vapour cell. In short, the negative group velocity observed in the experiments is due to the improper comparison of the advancement shift of the peak of the pulse with the time required for the peak of the pulse to pass through the cell, because the former also includes the contribution of pulse deformation. In the experiment, the wave front does not shift to an earlier time, its velocity is still $c$.

In fact, the concept of group velocity used to describe the average velocity of wave packet has its own problems. In non-dispersive medium, the propagation velocities of all the frequency components of the wave packet are equal, and the wave packet can propagate without deformation. In this case, the phase velocity is equal to the group velocity. In dispersive medium, the phase velocities of different frequency components are different, and the wave packet will be deformed. How to define the propagation velocity of a deformed wave packet becomes a key problem. Let's first see the definition of group velocity, which is $v_g = d\omega / dk$, where $\omega$ is angular frequency and $k = 2\pi / \lambda$ the wave number. As $\omega = v_p k$, we have

$$v_g = \frac{\mathrm{d}\omega}{\mathrm{d}k} = \frac{\mathrm{d}(v_p k)}{\mathrm{d}k} = v_p + k\frac{\mathrm{d}v_p}{\mathrm{d}k} = v_p - \lambda\frac{\mathrm{d}v_p}{\mathrm{d}\lambda} \tag{8}$$

which is just the expression derived by Rayleigh. It can be further transformed as

$$v_g = v_p - \lambda \frac{\mathrm{d}v_p}{\mathrm{d}\lambda} = \frac{c}{n} - \frac{c}{nf}\frac{\mathrm{d}(c/n)}{\mathrm{d}(c/nf)} = \frac{c}{n}\left(1 - \frac{f}{n}\frac{\mathrm{d}n}{\mathrm{d}f}\right) \tag{9}$$

On the other hand, starting from $k = \omega / v_p$, we have

$$v_g = \frac{\mathrm{d}\omega}{\mathrm{d}k} = \frac{\mathrm{d}\omega}{\mathrm{d}(\omega/v_p)} = \frac{\mathrm{d}\omega}{\frac{\mathrm{d}\omega}{v_p} - \frac{\omega}{v_p^2}\mathrm{d}v_p} = \frac{v_p}{1 - \frac{\omega}{v_p}\frac{\mathrm{d}v_p}{\mathrm{d}\omega}}$$

$$= \frac{c/n}{1 - \frac{f}{c/n}\frac{\mathrm{d}(c/n)}{\mathrm{d}f}} = \frac{c}{n}\frac{1}{1 + \frac{f}{n}\frac{\mathrm{d}n}{\mathrm{d}f}} \tag{10}$$

It can be seen that only when $|f\mathrm{d}n / n\mathrm{d}f| << 1$, Eqs. (9) and (10) are approximately equal, and the group velocity is meaningful. In the case of $|f\mathrm{d}n / n\mathrm{d}f| >> 1$, Eqs. (9) and (10) are far from each other, thus the definition of $v_g$ loses its meaning. For anomalously dispersive medium we have $f\mathrm{d}n/\mathrm{d}f < 0$. If $n + f\mathrm{d}n/\mathrm{d}f < 0$, negative group velocity will appear in (10), which is the theoretical foundation of the negative group velocity claimed by Wang and other authors. However, if we start from (9), we will get positive group velocity for anomalously dispersive medium in any case. In the experiment of Wang et al., the refractive index $n$ is measured using a radio-frequency interferometric technique and the group velocity index $n_g = n + f\mathrm{d}n/\mathrm{d}f = -330 \pm 30$ is derived. Since $n > 0$, we have $|f\mathrm{d}n/\mathrm{d}f| >> 1$. In this case, $v_g$ has already lost its meaning, so it cannot be inferred that the negative group velocity was observed in the experiment. To illustrate this point, we may increase the length of the cesium vapour cell. During the propagation process, there exists a limit for the deformation of the pulse (the advancement shift of the peak), which cannot exceed half of the whole pulse duration, as shown in Fig. 4. In other words, the peak of the pulse cannot arrive earlier than the wave front. Then when increasing the length of the cell gradually, we will observe that the propagation velocity of the peak of the pulse varies from negative group velocity to superluminal velocity, and finally becomes subluminal. However, it is difficult to build a large caesium vapour cell, we will propose an easy experiment to verify this idea in the next section.

Pulse reshaping can be used to explain many experiments, most of which are similar to that of Wang et al., so we will not discuss them. Now let's see a slightly different experiment. Budko found that there is a negative group velocity in the near region of the antenna [29]. The experimental setup is shown in Fig. 5, in which the distance between the receiving antenna and the transmitting antenna is less than 10 mm. From the previous analysis, we know that there exists evanescent wave in the experiment, but this kind of tunneling can only obtain superluminal velocity. In order to explain the negative group velocity observed in the experiment, we have to find other reason. When the two antennas are placed very close, the two antennas will form a coupling capacitor. In this case, there are not only radiation and evanescent wave components in the near region of the antenna, but also direct coupling component through the capacitor, and the coupling component is dominant. The circuit on the right side in Fig. 5 can be equivalent to an RC circuit, as shown in Fig. 6.

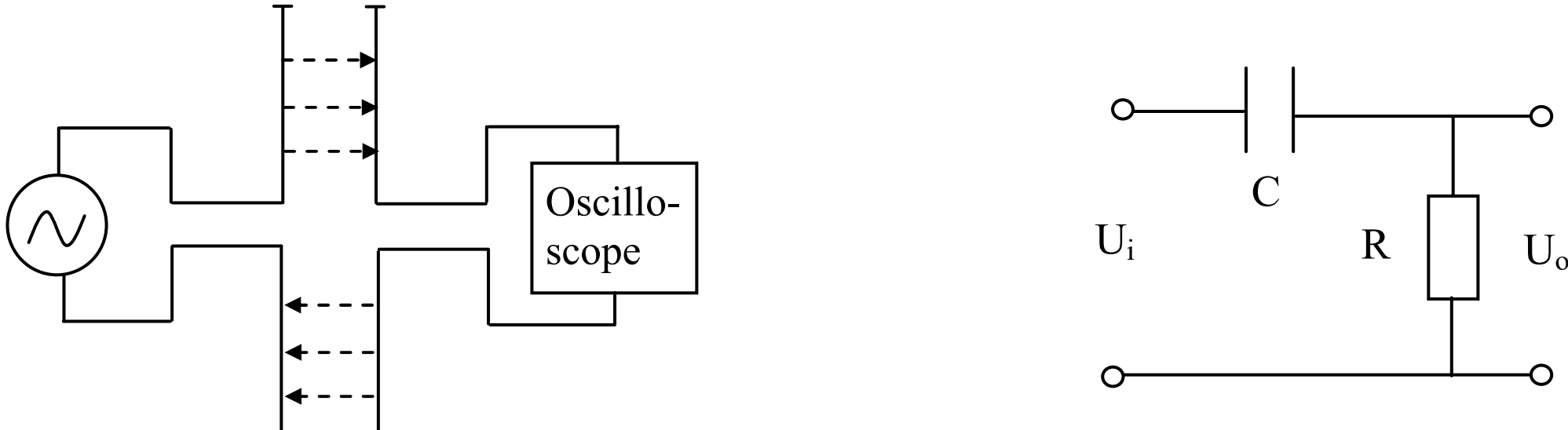

Fig. 5 The experimental setup of Budko.

Fig. 6 The equivalent circuit of Fig. 5.

The voltage $U_o$ of the load $R$ is

$$U_o = \frac{R}{\frac{1}{i\omega C} + R} U_i = \left(\frac{\omega^2 R^2 C^2}{1+\omega^2 R^2 C^2} + i\frac{\omega RC}{1+\omega^2 R^2 C^2}\right) U_i \tag{11}$$

The phase is

$$\phi = \tan^{-1}\frac{1}{\omega RC} \tag{12}$$

The group delay is

$$t = -\frac{\mathrm{d}\phi}{\mathrm{d}\omega} = \frac{RC}{1+\omega^2 R^2 C^2} \tag{13}$$

We see from Eq. (13) that as the frequency $\omega$ of the signal increases, the transmission time through the load $R$ decreases, which agrees with the characteristic of anomalous dispersion. Thus the experimental setup of Fig. 5 can indeed obtain negative group velocity. In fact, negative group velocity and superluminal velocity with RC circuit has been observed in Ref. [30], and a RLC bandpass amplifier has been used to demonstrate negative group delay in Ref. [31].

## 3 Proposed optical experiments

### 3.1 Experimental verification of the spatial extension of a single photon wave packet

If we suppose that a single photon is a wave packet occupying certain spatial volume, it's easy to understand the single photon double-slit interference experiment, i.e. one half of the wave packet passes through one slit and the other half through the other slit. Now we test this hypothesis. We do the experiment based on the experimental setup of Aspect et al. [32], as shown in Fig. 7, where the single photon was produced by the cascade radiation in calcium. We first verify the fact that the interference outputs of $D_1$ and $D_2$ will disappear when the optical path length difference is larger than the coherence length of the photon. We then increase the distance L between the beam splitter (BS) and the reflecting mirror $M_2$. We expect that when L is larger than the transverse extension of the photon wave packet (which is equivalent to increasing the width of double-slit in interference experiment), the interference will also disappear. In fact, such results might be conceived even if we don't do the experiment. What is important is that it gives us the clear evidences that a single photon has certain spatial extension.

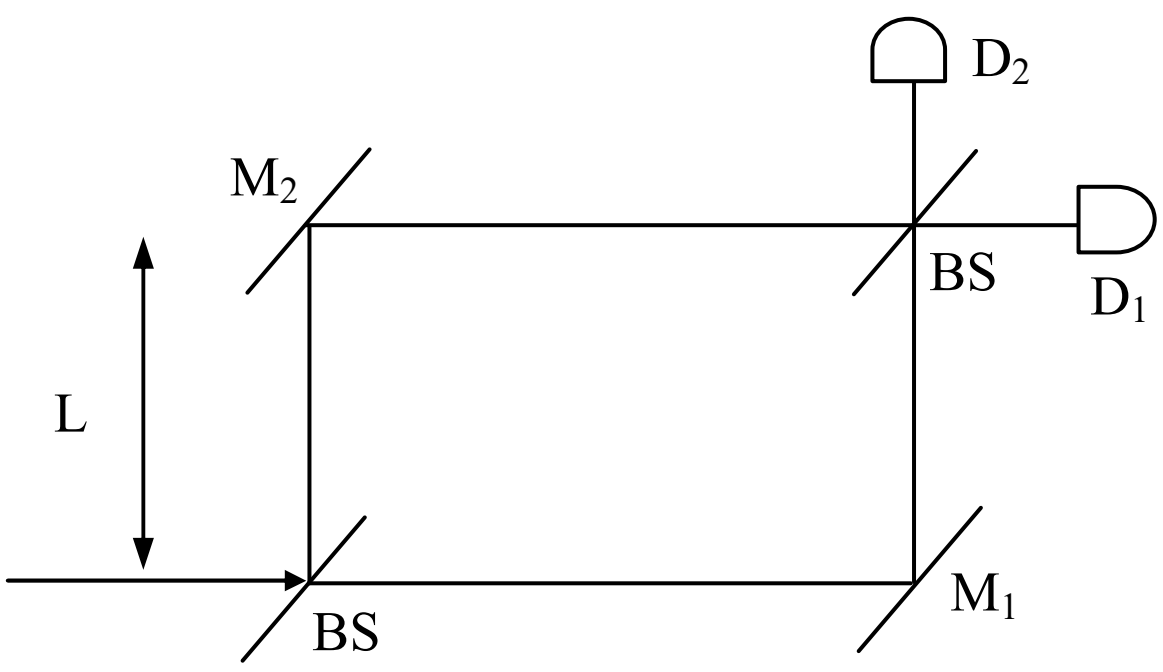

Fig. 7 Experimental test of spatial extension of a single photon.

### 3.2 Experimental test of superluminal propagation within coherence length

According to our hypothesis, no photons can tunnel through a barrier if the coherence length is smaller than twice the barrier thickness. While in terms of evanescent wave theory, the tunneling probability of photons decreases exponentially with tunneling distance but will not be zero at large distances. In order to test which theory is correct, we shall carry out the following experiment. Let a beam of microwave or light pulse be incident upon an undersized waveguide or a double-prism, as shown in Fig. 8. We then gradually increase the thickness of the barrier and measure the power of the tunneled microwave to see whether it decreases exponentially with tunneling distance or the signal is undetectable beyond a certain distance no matter how we increase the power radiated. If it turns out that the latter situation is correct, then our hypothesis is correct.

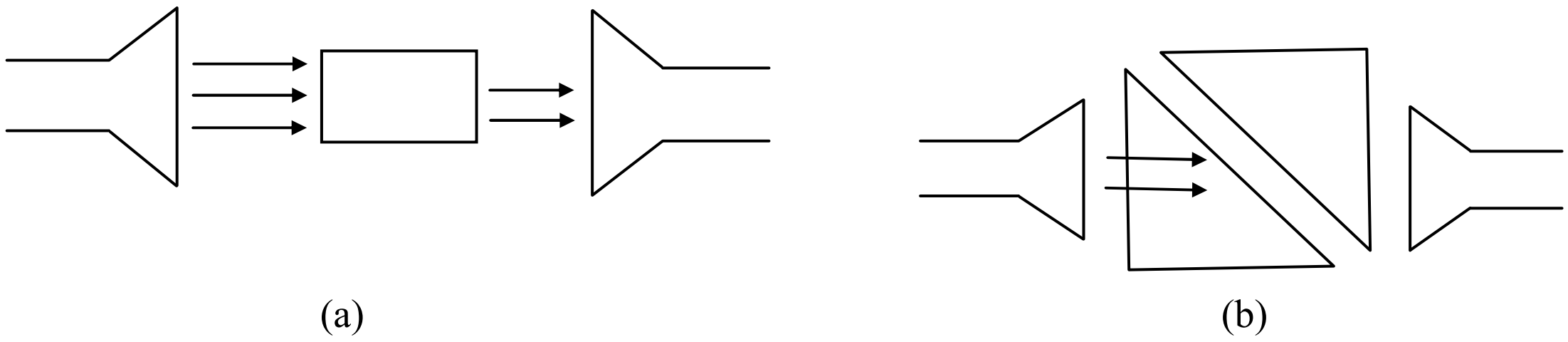

(a) (b)

Fig. 8 Microwave tunnels through undersized waveguide (a) or double-prism (b).

### 3.3 Experimental test of subluminal propagation in anomalously dispersive medium

We have supposed that the superluminal phenomena in anomalously dispersive medium are the consequence of reshaping of the pulse, and superluminal propagation will disappear at large distances. As the experiment of Wang et al. is difficult to realize on a large scale, while it's easy to realize by using tunneling of electric pulse through a coaxial photonic crystal, we adopt the methods in [33] and [34], where the experimental setups are similar. The main difference is the use of a coaxial photonic crystal in [34] with a higher impedance mismatch that permits access to a negative group velocity of -1.2 $c$, while the experiment in [33] only obtained the results of 2~3.5 $c$. In terms of our theory, when the length of the coaxial cable increases, the propagation velocity of the peak of the pulse will vary from negative group velocity to superluminal velocity, and finally becomes subluminal. Suppose we adopt the experimental setup of [33], whose simplified sketch is shown in Fig. 9. One output of the signal generator is connected directly to the oscilloscope as the reference signal, and the other output passes through a photonic crystal made of alternating quarter-wavelength segments of two different impedance coaxial cables. Now we estimate the length $L$ required for the photonic crystal to obtain subluminal velocity of the electric pulse. The phase velocity of the electric pulse is $0.66c$ in [33]. Suppose the peak of the pulse can shift at most earlier to the wave front. For a pulse with duration of 2us adopted in the experiment of [33], we have

$$\frac{L}{0.66c}-\frac{L}{c}\geq 10^{-6}. \tag{14}$$

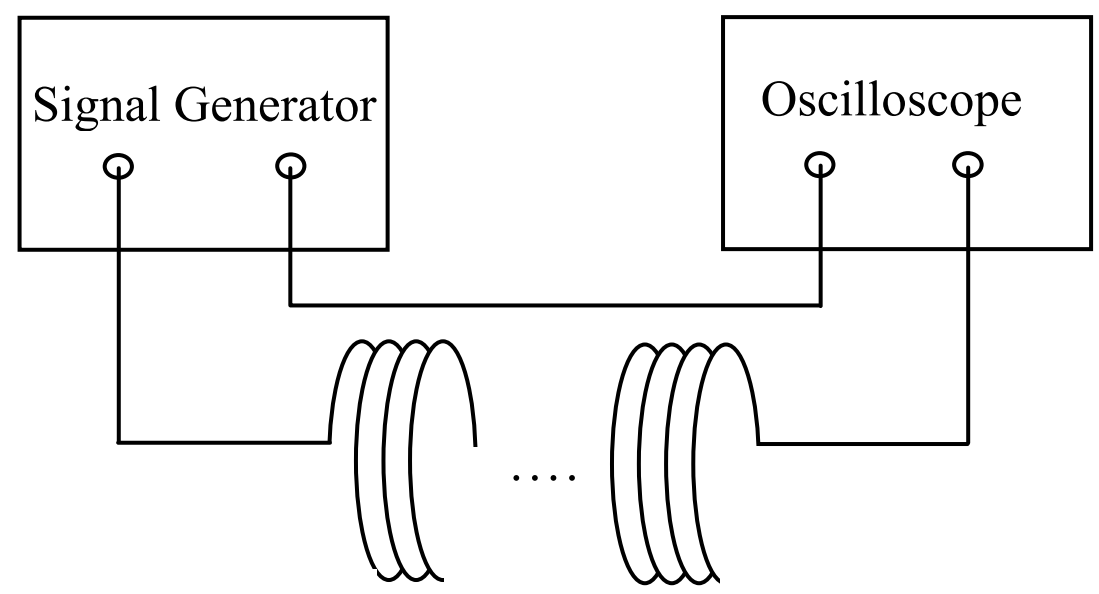

Fig. 9 Experimental test of subluminal propagation of electric pulse at large distances.

It follows $L\geq 600\,\text{m}$. By contrast, the length of the coaxial cable used in [33,34] is 120m. In fact, the peak of the pulse cannot shift to the wave front, so we can observe subluminal propagation at a length less than 600m.

## 4 Conclusion

When light propagates as a wave train in vacuum or medium, its velocity cannot be superluminal. In the presence of a barrier, a few of the photons can tunnel through the barrier as particles (evanescent wave) and their velocities can exceed $c$. But this superluminal motion can only last within the coherence length and a short time. For a particle with non-vanishing mass, its velocity cannot be superluminal whether in wave or particle state due to the uncertainty principle. Superluminal propagation is always accompanied with attenuation and distortion of the signal. As for the superluminal propagation in anomalously dispersive medium, we may regard it as the consequence of reshaping of the pulse. At large distances the propagation velocity of the pulse will be subluminal. So it's not practical to realize superluminal communication with existing experimental devices. Both of the two superluminal effects are not at odds with causality. The proposed experiments may be used to further test superluminal phenomena.